\documentclass[aps,preprint,nofootinbib,preprintnumbers,showpacs]{revtex4}
\usepackage{amsmath,amssymb}
\usepackage{bm}
\usepackage{comment}
\usepackage{graphicx}
\usepackage{pdfpages,color}

\renewcommand\d{\partial}
\newcommand\x{\mathbf{x}}

\newcommand\<{\langle}
\renewcommand\>{\rangle}
\renewcommand\Im{\mathop{\mathrm{Im}}}

\preprint{EFI-13-22}
\begin{document}
\title{Spectral Sum Rules and Magneto-Roton as Emergent Graviton in
  Fractional Quantum Hall Effect} 
\author{Siavash Golkar, Dung X. Nguyen, and Dam T.~Son} 
\affiliation{Enrico Fermi Institute, James Franck Institute and Department of Physics, University of Chicago, Chicago, Illinois 60637, USA}

\begin{abstract}
We consider gapped fractional quantum Hall states on the lowest Landau
level when the Coulomb energy is much smaller than the cyclotron
energy.  We introduce two spectral densities, $\rho_T(\omega)$ and
$\bar\rho_T(\omega)$, which are proportional to the probabilities of
absorption of circularly polarized gravitons by the quantum Hall
system.  We prove three sum rules relating these spectral densities
with the shift~$\mathcal S$, the $q^4$ coefficient of the static
structure factor $S_4$, and the high-frequency shear modulus of the
ground state $\mu_\infty$, which is precisely defined.  We confirm an
inequality, first suggested by Haldane, that $S_4$ is bounded from
below by $|\mathcal S-1|/8$.  The Laughlin wavefunction saturates this
bound, which we argue to imply that systems with ground state wavefunctions close
to Laughlin's absorb gravitons of predominantly one circular
polarization.  We consider a nonlinear model where the sum rules are
saturated by a single magneto-roton mode.  In this model, the
magneto-roton arises from the mixing between oscillations of an
internal metric and the hydrodynamic motion.  Implications for
experiments are briefly discussed.
\end{abstract}

\maketitle

\section{Introduction}

Fractional quantum Hall (FQH) systems represent the paradigm for
interacting topological states of matter.  Much attention has been
concentrated on the topological properties of the quantum Hall states,
encoded in the ground state wavefunction and the statistics of
quasiparticle excitations.  On the other hand, FQH systems also have a
neutral collective excitation known as the magneto-roton.  The magneto-roton
is an excitation within one Landau level and hence has energy at the
interaction (Coulomb) energy scale.  The existence of the
magneto-roton was suggested in the classic work of Girvin, MacDonald,
and Platzman~\cite{Girvin:1986zz}, in which Feynman's theory of the
roton in superfluid helium is extended to the FQH case.  In this
picture the magneto-roton is visualized as a long-wavelength density
fluctuation.  Within the composite fermion theory, the magneto-roton
is interpreted as a particle-hole bound state, in which the particle
lies in an empty composite-fermion Landau level, and the hole lies in
a filled one.  In the composite-boson approach, the magneto-roton is
interpreted as vortex-antivortex bound state.  The magneto-roton has
been observed in Raman scattering
experiments~\cite{Pinczuk,Pinczuk:2000} and more recently, in
experiments using surface acoustic waves~\cite{Kukushkin}.  At nonzero
wavenumbers, there is some evidence in favor of more than one
magneto-roton mode~\cite{Pinczuk:2005}.

Very recently, Haldane has proposed a drastically different
interpretation of the
magneto-roton~\cite{Haldane:2009,Haldane:2011,Haldane-model-wf}.  He
argues that there is a dynamic degree of freedom in FQH systems which
can be interpreted as an internal metric.  In this picture, the
magneto-roton at long wavelength is the quantum of the fluctuations of
this metric.

In this paper, we derive some new results related to the physics of a
gapped FQH system at the lowest Landau level ($\nu<1$) at the
interaction energy scale.  We assume that the Coulomb energy
is much smaller than the cyclotron energy.  First, we derive new,
exact sum rules involving the spectral densities of the traceless part
of the stress tensor.  The two components of the traceless part of the
stress tensor are $T_{zz}=\frac14(T_{xx}-T_{yy}-2iT_{xy})$, $T_{\bar
  z\bar z}=\frac14(T_{xx}-T_{yy}+2iT_{xy})$ (here $z=x+iy$), hence we
can define two spectral densities:
\begin{align}
  \rho_T(\omega) &= \frac1N \sum_n\big|\< n | \int\!d^2\!x\, T_{zz}({\bf x}) |0 \>\big|^2
     \delta(\omega-E_n), \label{rho}\\
  \bar\rho_T(\omega) &= \frac1N \sum_n|\big\< n | \int\!d^2\!x\, 
     T_{\bar z\bar z}({\bf x}) |0 \>\big|^2
     \delta(\omega-E_n), \label{rhobar}
\end{align}
where $N$ is the total number of particles in the system, $|0\rangle$
is the ground state and the sums are taken over all excited states
$|n\rangle$.  Physically, these spectral densities are proportional to
the probability that a circularly polarized graviton with energy
$\omega$ falling perpendicularly to the system is absorbed.  The two
functions correspond to the two circular polarizations of the
graviton.  Without a complete solution to the FQH problem, we do not know
$\rho_T(\omega)$ and $\bar\rho_T(\omega)$, but
for a gapped FQH system we expect these functions to be zero below a gap
$\Delta_0$ and to fall to zero when
$\omega$ increases far above $\Delta_0$.  If there is a well-defined magneto-roton at
$q=0$, we expect it to appear as peaks in the spectral densities.

We will show that the spectral densities satisfy three sum rules,
\begin{align}
  \int_0^\infty\!\frac{d\omega}{\omega^2} 
     [\rho_T(\omega) - \bar \rho_T(\omega)]& = \frac{\mathcal S-1}8\,,
     \label{shift-sr-intro}\\
  \int\limits_0^\infty\!\frac{d\omega}{\omega^2}\, 
  [\rho_T(\omega) + \bar \rho_T(\omega)]   &= S_4,
     \label{S4-sr-intro}\\
  \int_0^\infty\!\frac{d\omega}{\omega}
   [\rho_T(\omega) + \bar \rho_T(\omega)]   &= \frac{\mu_\infty}{\rho_0}\,.
   \label{elasticity-sr-intro}
\end{align}
In Eq.~\eqref{shift-sr-intro}, $\mathcal S$ is the shift of the ground
state, defined as the offset in the relationship between the number of
magnetic flux quanta $N_\phi$, the filling factor $\nu$ and the number
of electrons $Q$ when the latter are put on a sphere:
$Q=\nu(N_\phi+\cal S)$~\cite{Wen:1992ej}.  In Eq.~\eqref{S4-sr-intro},
$S_4$ is the coefficient governing the low-momentum behavior of the
projected structure factor~\cite{Girvin:1986zz}: $\bar
s(q)=S_4(q\ell_B)^4$, where $\ell_B$ is the magnetic length.  Finally,
in Eq.~\eqref{elasticity-sr-intro}, $\mu_\infty$ is the high-frequency
elastic modulus, which will be defined exactly later in the text [see
  Eqs.~\eqref{falpha} and \eqref{muinf-def}], and $\rho_0$ is the
particle number density in the ground state.  In all sum rules, the
limit $m\to0$ of the spectral densities is taken first, before the
upper limit of integration is taken to infinity.  In this order of
limits, the integral in each sum rule is dominated by $\omega$ of
the order of the Coulomb energy.

The sum rule~\eqref{shift-sr-intro} is particularly interesting, as it
establishes a connection between a topological characteristic of the
ground state (the shift) and dynamic information (the spectral
densities).  In the $\nu=1$ integer quantum Hall state, the sum rule
becomes trivial, as there is no degree of freedom at the interaction
energy scale (hence $\rho_T=\bar\rho_T=0$), and the shift is $\mathcal
S=1$, so both sides of the sum rule vanish.

Using the three sum rules, we derive some inequalities between
different observables in the FQH states.  One of these inequalities,
previously derived by Haldane~\cite{Haldane:2009}, places a lower
bound on the coefficient of the $q^4$ asymptotics of the projected
static structure factor, which is saturated by the Laughlin's trial
wavefunction.

We will also consider a simple model where the sum rules are saturated
by one magneto-roton mode, which manifests as the oscillation of the
internal metric of the fluid mixed with the hydrodynamic fluid motion.
The model provides a concrete realization of Haldane's idea of an
internal metric degree of freedom in FQH systems~\cite{Haldane:2009}.
This model is not meant to be exact, however it does exhibit some of
the characteristic properties of the observed magneto-roton modes.  We
discuss the polarization properties of the magneto-roton in this
model, which may be measurable in future experiments.

\section{The effective action}

\subsection{Review of the Newton-Cartan formalism}

Recently, one of the authors has proposed the use of nonrelativistic
general coordinate invariance as a way to constrain the dynamics of
quantum Hall systems~\cite{Son:2013rqa} (for related work, see
Refs.~\cite{Wiegmann,Abanov}).  Although the method can be thought of
as ``gauging'' the Galilean invariance, the local symmetry remains
nontrivial in the limit of zero bare electron mass, and hence is a
symmetry intrinsic to the physics of electrons at the lowest Landau
level, coupled to electromagnetism and gravity.

In Ref.~\cite{Son:2013rqa}, the attention was focused on the regime of
long wavelengths (much larger than the magnetic length) and low
frequencies (much smaller than the gap).  In this paper we will relax
the latter condition, allowing for energies comparable to the gap.  In
this regime, terms with arbitrary number of time derivatives must be
taken into account in the effective action. However, as we will
demonstrate, we can still obtain nontrivial relationships by expanding
in the number of spatial derivatives.

We briefly review the main result of Ref.~\cite{Son:2013rqa} here.
The effective Lagrangian describing the response of a gapped quantum
Hall state to external electromagnetic ($A_0$, $A_i$) and gravitational
perturbations ($h_{ij}$), in the massless limit, is:
\begin{equation}\label{L-NC}
  \mathcal L = \frac\nu{4\pi}\varepsilon^{\mu\nu\lambda} a_\mu\d_\nu
  a_\lambda - \rho v^\mu(\d_\mu\varphi-\tilde A_\mu -s\omega_\mu +a_\mu) 
  + {\mathcal L}_0[\rho, v^i, h_{ij}].
\end{equation} 
Here $v^\mu=(1,v^i)$; $a_\mu$, $\rho$ and $v^i$ are dynamical fields, with
respect to which one should extremize the action; and $\tilde A_\mu$ is related
to the external electromagnetic potential by:
\begin{equation}
  \tilde A_0 = A_0 - \frac12\varepsilon^{ij}\d_i(h_{jk}v^k), \qquad
  \tilde A_i = A_i,
\end{equation}
$\omega_\mu$ is the spin connection of the Newton-Cartan space
$(h_{ij}, v^i)$, defined through the derivatives of the vielbein
$e^a_i$ ($h_{ij}=e^a_i e^a_j$):
\begin{equation}
  \omega_ 0 = \frac12 \epsilon^{ab} e^{aj}\d_0 e^b_j 
              + \frac12 \varepsilon^{ij}\d_i(h_{jk}v^k), \qquad
  \omega_i = \frac12 \epsilon^{ab} e^{aj}\nabla_{\!i} e^b_j.
\end{equation}
Finally, $\mathcal L_0$ contains all ``non-universal'' terms, i.e., terms that cannot be fixed by symmetry arguments alone.

There are two parameters that enter the Lagrangian~\eqref{L-NC}:
$\nu$, which is identified with the filling factor, and $s$,
identified with the orbital spin (per particle) and is related to the
shift by $s=\mathcal S/2$.  In Ref.~\cite{Son:2013rqa} it was found
that these two parameters control some quantities, most notably the
$q^2$ part of the Hall conductivity at zero frequency ($q$ being the
wavenumber of the perturbation).

\subsection{Physics at the Coulomb energy scale}

It was also found in Ref.~\cite{Son:2013rqa} that most physical
quantities, e.g., the same $q^2$ part of the Hall conductivity, but
calculated at nonzero frequency, are not fixed by $\nu$ and $s$ alone.
The same is true for the $q^4$ term in the density-density correlation
function.  Physically, these quantities depend crucially on the
physics happening at the Coulomb energy scale~$\Delta$.  The physics
of the gapped excitations is contained in the non-universal part of
the Lagrangian $\mathcal L_0$.

Terms in $\mathcal L_0$ can be organized in a series over powers of
derivatives.  We will be interested in the physics at long
wavelengths, $q\ell_B\ll1$.
The expansion parameter in frequency would be $\omega/\Delta$,
however since we are interested in physical phenomena at the scale
$\Delta$, we need to keep terms to all orders in time derivatives.

A consistent power counting scheme is to consider fluctuations of the
metric $h_{ij}$ and the gauge potentials $A_0$, $A_i$ as $O(1)$, and
expand in powers of the spatial derivatives.  In this work, we will be
interested only in the response of the quantum Hall systems to
\emph{unimodular} metric perturbations, i.e., those in which the
perturbed metric $h_{ij}$ has determinant equal to one, and to
perturbations of the scalar potential $A_0$, i.e., perturbations
corresponding to a longitudinal electric field, without changing the
magnetic field.  In this case, the lowest non-trivial terms entering
$\mathcal L_0$ are  $O(q^4)$.  These we parameterize, without loss of 
generality\footnote{As an example, the term $(\d_i \rho)^2$ is of
higher order, because fluctuations of $\rho$ are of order $q^4$, as evident
from Eq.~(\ref{sbar}) below.  The same is true for $(\d_i v^i)^2$: 
because of charge
conservation $\d_i v^i \sim \d_t\rho \sim q^4$.}, with two functions
$F(\omega)$ and $G(\omega)$.  In the following equation the frequency
$\omega$ is replaced by $iv\cdot\nabla$ where $\nabla$ is the 
Newton-Cartan covariant derivative~\cite{Son:2013rqa},
\begin{equation}\label{L0-FG}
  \mathcal L_0 = -\frac\rho 4 \left[\sigma^{\mu\nu} F\big(iv\cdot\nabla\big)
    \sigma_{\mu\nu}  + \tilde\sigma^{\mu\nu}G\big(iv\cdot\nabla\big) (v\cdot\nabla
    \sigma_{\mu\nu})
  \right].
\end{equation}
Here $\sigma_{\mu\nu}$ is the traceless part of shear tensor 
(see \cite{Son:2013rqa} for precise definition) and
$\tilde\sigma^{\mu\nu}$ is defined as:
\begin{equation}
  \tilde\sigma^{\mu\nu}=
  \frac12 (\varepsilon^{\mu\alpha\gamma} h^{\nu\beta} 
    + \varepsilon^{\nu\alpha\gamma} h^{\mu\beta}) \sigma_{\alpha\beta}
    n_\gamma,
\end{equation}
where $n_\mu=(1,\mathbf 0)$.

We will work only to quadratic order, hence we only need to know
the leading terms in the spatial components of the shear tensor:
\begin{equation}
  \sigma_{ij} = \d_i v_j + \d_j v_i + \dot h_{ij}
    - \delta_{ij} (\d_k v_k + \frac12 \dot h).
\end{equation}
The quadratic part of the Lagrangian is then:
\begin{equation}\label{L0-quad}
  \mathcal L_0 = -\frac{\rho_0}4\left[ \sigma_{ij} F(i\d_t) \sigma_{ij}
  + \tilde \sigma_{ij} G(i\d_t) \dot\sigma_{ij} \right].
\end{equation}

Here, for any symmetric traceless tensors $A_{ij}$ we define
$\tilde A_{ij} = \frac12 ( \epsilon_{ik}A_{kj} + \epsilon_{jk}A_{ki} )$
which is again a symmetric traceless tensor.  It is also easy to show
${\tilde {\tilde A}}_{ij} = - A_{ij}$ and
$\tilde A_{ij} B_{ij} = - A_{ij}\tilde B_{ij}$.

\subsection{Gravitational response, spectral representations and shift 
sum rule}

We now relate the two functions $F(\omega)$ and $G(\omega)$ to the
spectral densities of the stress tensor $\rho_T$ and $\bar\rho_T$.  The two-point function of
the stress tensor can be read directly from the action, and is
simplest for the traceless components at zero spatial momentum.  After
a simple calculation we get:
\begin{align}
  \< \bar T T\>_\omega &= -\frac\omega 4 s\rho_0 + \frac{\omega^3}2 \rho_0 G 
      + \frac{\omega^2}2 \rho_0 F,\\
  \< T\bar T\>_\omega &= \phantom{+}\frac\omega 4 s\rho_0
      - \frac{\omega^3}2 \rho_0 G 
      + \frac{\omega^2}2 \rho_0 F.
\end{align}
The spectral densities defined in Eqs.~\eqref{rho} and \eqref{rhobar}
are related to $F$ and $G$ by:
\begin{align}
  \rho_T(\omega) &= -\frac{\omega^2}{2\pi}\Im (F+\omega G),\\
  \bar \rho_T(\omega) &= 
    -\frac{\omega^2}{2\pi} \Im (F-\omega G),
\end{align}
where we have extended the definition of the spectral densities to
negative $\omega$'s by requiring $\rho_T(-\omega)=\bar\rho_T(\omega)$.
The functions $F(\omega)$ and $G(\omega)$ are regular in the
$\omega\to0$ limits, and as we shall explain below [see Eqs.~(\ref{omega2G}) and (\ref{omega2F})] they should both fall as
$1/\omega^2$ when $\omega\gg\Delta$.  From these behaviors we can
write down the spectral representations of $F$ and $G$:
\begin{align}
  F(\omega) &= 2 \int\limits_0^\infty\!\frac{d\omega'}{\omega'} \, 
    \frac{\rho_T(\omega') + \bar\rho_T(\omega')}{\omega^2-\omega'^2+i\epsilon},
  \label{F-disp}
   \\
  G(\omega) &= 2 \int\limits_0^\infty\!\frac{d\omega'}{\omega'^2} \,
    \frac{\rho_T(\omega')-\bar\rho_T(\omega')}{\omega^2-\omega'^2+i\epsilon}. 
  \label{G-disp}
\end{align}


The Hall viscosity, as a function of frequency, can be related through a Kubo's
formula to the parity-odd part of the two-point function of the stress
tensor~\cite{Bradlyn:2012ea}.  We find:
\begin{equation}
  \eta_{\rm H}(\omega) = \rho_0\left( \frac s2 - \omega^2 G(\omega)\right).
\end{equation}
At $\omega\to0$ this equation gives the relationship between the
(zero-frequency) Hall viscosity and the shift: $\eta_{\rm
  H}(0)=\rho_0\mathcal S/4$.  At frequencies much larger than $\Delta$,
interactions can be neglected and the Hall viscosity is determined
completely by the Berry phase of each orbital under homogeneous metric
deformation.  The computation of the high-frequency Hall viscosity 
(where ``high'' means frequencies much larger than the Coulomb energy 
scale, but still much smaller than the cyclotron energy)
proceeds in exactly the same way as in the integer quantum Hall 
case~\cite{Avron:1995fg}, and the result is
$\eta_\mathrm{H}(\infty)=\rho_0/4$.  Thus we find:
\begin{equation}\label{omega2G}
\lim_{\omega\to\infty}\omega^2 G(\omega) = \frac{\mathcal S-1}4,
\end{equation}
and using Eq.~(\ref{G-disp}) we derive our first sum rule:
\begin{equation}\label{shift-sr}
  \int_0^\infty\!\frac{d\omega}{\omega^2} 
     [\rho_T(\omega) - \bar \rho_T(\omega)] = \frac{\mathcal S-1}8.
\end{equation}

\subsection{Static structure factor and high-frequency shear modulus}
\label{sec:shear_mod}

We now derive two sum rules involving the sum of the two spectral
densities $\rho_T(\omega)+\bar\rho_T(\omega)$.  Computing the
two-point function of the density from the action from~\eqref{L-NC}
and \eqref{L0-quad}, we find:
\begin{equation}\label{sbar}
  \int\!d^3x\, e^{i\omega t - i{\bf q}\cdot {\bf x}} 
      \< T \rho(t,\x) \rho(0,{\bf 0})\> = i\rho_0 (q\ell_B)^4 F(\omega),
\end{equation}
Integrating both sides over $\omega$, we get:
\begin{equation}\label{sbar1}
  \bar s(q)
  = i(q\ell_B)^4\!\! \int\limits_{-\infty}^\infty\!
    \frac{d\omega}{2\pi} \, F(\omega),
\end{equation}
where $\bar s(q)$ is the projected static structure
factor~\cite{Girvin:1986zz}.  The reason we get the projected
structure factor instead of the unprojected one is that we are working
in the limit of zero band mass, and so when we took the integral over
$\omega$, implicitly we have assumed that upper limit of integration is
still much smaller than the cyclotron frequency $B/m$.  
By using
Eq.~(\ref{F-disp}), we find the second sum rule:
\begin{equation}\label{S4-sr}
  \int\limits_0^\infty\!\frac{d\omega}{\omega^2}\, 
  [\rho_T(\omega) + \bar \rho_T(\omega)]   = S_4,
\end{equation}
where $S_4=\lim\limits_{q\to0}\bar s(q)/(q\ell_B)^4$.

The third sum rule again comes from the stress response at large
$\omega$.  For this purpose, it is convenient to describe the motion
of the fluid in terms of the displacement $u^i$, which is related
to the velocity by $v^i=\dot u^i$.  The $F$ term in the action now reads:
\begin{equation}
  - \frac{\rho_0}4\int\!d^2x\, \omega^2 F(\omega) 
  \left[\d_i u_j + \d_j u_i+h_{ij}- \delta_{ij} 
  \left(\d\cdot u + \frac h2\right)\right]^2.
\end{equation}
In the limit $\omega\to\infty$, the $G$ term, having an extra time
derivative, does not contribute to the energy. Hence, we are left
with only the $F$ contribution above, which takes the exact same 
form as the the deformation energy of
a solid, with the shear modulus $\mu_\infty$ given by:
\begin{equation}\label{omega2F}
 \mu_\infty = \frac{\rho_0}2  
 \lim_{\omega\to\infty} \omega^2F(\omega) .
\end{equation}
Using the spectral representation of $F(\omega)$ in \eqref{F-disp}, we find:
\begin{equation}\label{elasticity-sr}
  \int_0^\infty\!\frac{d\omega}{\omega}
   [\rho_T(\omega) + \bar \rho_T(\omega)]   = \frac{\mu_\infty}{\rho_0}\,.
\end{equation}

The high-frequency shear modulus $\mu_\infty$ was introduced in the
phenomenological model of 
Refs.~\cite{Tokatly:2006,TokatlyVignale:2007}.  We now give the
precise meaning of this constant.  From our discussion, we know that
$\mu_\infty$ characterizes the stress
response of the system under uniform metric perturbations with
frequencies much higher than the Coulomb energy scale, but much lower
than the cyclotron energy.  Since at these
frequencies the Coulomb interaction between electrons can be ignored,
each particle evolves independently under such a perturbation.  The
orbital of each electron is continuously deformed and projected down to the lowest Landau level.  In this way, we can completely determine
the wavefunction of the deformed state from that of the the ground
state.  For example, consider a metric perturbation in which
the $x$ coordinate is stretched by a factor of $e^{\alpha/2}$ while the
$y$ coordinate compressed by $e^{-\alpha/2}$.  If we denote the ground
state wave function as:
\begin{equation}
  \Psi(z_i) = f(z_i) \exp\Bigl(-\sum_i |z_i|^2/4\ell_B^2\Bigr),
\end{equation}
the deformed state $|\Psi_\alpha\>$ is obtained by replacing $f(z_i)$ by
$f_\alpha(z_i)$,
\begin{equation}\label{falpha}
  \Psi_\alpha(z_i) = f_\alpha(z_i) \exp\Bigl(-\sum_i |z_i|^2/4\ell_B^2\Bigr),
  \qquad
  f_\alpha(z_i)= \exp\Bigl[ \frac\alpha2 \sum_i 
  \Bigl( \ell_B^2  \frac{\d^2}{\d z_i^2} 
  - \frac{z_i^2}{4\ell_B^2}\Bigr)\Bigr] f(z_i).
\end{equation}
If we now use Laughlin's wavefunction to substitute for $f(z_i)$
above, the deformed states that we obtain coincide exactly with the
ones recently considered in Ref.~\cite{Haldane-Yang}.

The energy of these states is a function of $\alpha$ with minimum at
$\alpha=0$, and the high-frequency shear modulus is simply the
curvature of this function at the minimum:
\begin{equation}\label{muinf-def}
  \mu_\infty = \frac1A \frac{\d^2}{\d\alpha^2} \< \Psi_\alpha|\hat H
   | \Psi_\alpha\>|_{\alpha=0}.
\end{equation}
where $A$ is the total area of the system.
Equations~(\ref{muinf-def}) and (\ref{falpha}) define the constant
$\mu_\infty$ appearing in the sum rule~(\ref{elasticity-sr}).

\subsection{Inequalities following from the sum rules}

The sum rules have important
implications.  First, since $\rho_T$ and $\bar\rho_T$ are non-negative
spectral densities, comparing eqs.~\eqref{shift-sr} and \eqref{S4-sr}, we obtain the following inequality between $S_4$
and $\mathcal S$:
\begin{equation}\label{S4-lowerbound}
  S_4 \ge \frac{|\mathcal S-1|}8 \,.
\end{equation}
This inequality has been previously derived by
Haldane~\cite{Haldane:2011,Haldane-selfdual}.  For Laughlin's
fractions~$\nu=$~$1/(2p+1)$, $\mathcal S = 1/\nu$, and the inequality
becomes $S_4\ge (1-\nu)/8\nu$.  Remarkably, the Laughlin
wavefunction has $S_4$ saturating the lower bound.  Hence, if the
Laughlin wavefunction was the true wavefunction of the ground state,
that would imply $\bar\rho_T=0$ for all~$\omega$.

Read and Rezayi~\cite{ReadRezayi:2011} argued that the
inequality~(\ref{S4-lowerbound}) is actually an equality for all
lowest-Landau-level ground states with rotational invariance.  From
our derivation, we do not expect the equality to hold generally: the
spectral density $\bar\rho_T$ need not necessarily vanish. 
Nevertheless, the Laughlin wavefunction seems to be a very good
approximation to the true wavefunction of the Coulomb potential, thus
it is possible that for the true ground state of the Coulomb problem,
$\bar\rho_T$ is numerically much smaller than $\rho_T$.

Finally, we can also put a lower bound on the energy gap
$\Delta_0$ at $q=0$.  The energy gap may correspond not to a single
quasiparticle, but, for example, to a pair of magneto-rotons, in which
case $\Delta_0$ is the start of a continuum.  The inequality that
follows from comparing the sum rules \eqref{S4-sr} and \eqref{elasticity-sr} is:
\begin{equation}\label{ineq2}
  \Delta_0 \le \frac{\mu_\infty}{\rho_0 S_4},
\end{equation}
where equality would be achieved only when $\rho_T$ and $\bar\rho_T$
are proportional to $\delta(\omega-\Delta_0)$.  The equality in this
case has the same form as the Girvin-MacDonald-Platzman variational
formula for the magneto-roton energy, but in contrast to the latter,
both the numerator and the denominator in our formula are finite in
the limit $q\to0$.

By combining these two inequalities we can also write:
\begin{equation}\label{ineq3}
  \Delta_0 \le \frac{8\mu_\infty}{\rho_0 |\mathcal S-1|}, 
\end{equation}
which saturates under the conditions $\bar\rho_T=0$ and
$\rho_T(\omega)\sim\delta(\omega-\Delta_0)$.


\section{A gravitational model of the magneto-roton}

We now present a simple model where the sum rules are satisfied by
construction and are dominated by one single mode which is identified
with the magneto-roton.  To start, we adapt the Lagrangian
formulation of fluid dynamics~\cite{Dubovsky:2005xd}, in which the degrees 
of freedom of the
quantum Hall fluids are the Lagrangian coordinates $X^I(t,\x)$,
$I=1,2$.  The density and velocity of the fluid are given by:
\begin{equation}
  \rho v^\mu=\rho_0\varepsilon^{\mu\nu\lambda}\epsilon_{IJ}\d_\nu
X^I\d_\lambda X^J,
\end{equation}
such that the divergence of the current vanishes identically.  
The theory is required to
be invariant under volume-preserving diffeomorphisms in the $X^I$
space.  Imposing this condition sets the shape modulus to zero, thereby
ensuring that our action describes a fluid and not a solid.

The degree of freedom saturating the sum rules, is assumed to be a
unimodular metric tensor $G_{I\!J}$.  Physically, one should think of $G_{I\!J}$
as parameterizing the anisotropic deformation of the ground state, as
constructed in section \ref{sec:shear_mod}.  In this model, 
$G_{I\!J}$ is the only dynamical degree of freedom at the Coulomb
energy scale. 

The theory can either be written in $x$ space, treating $X^I$ as functions of
$t$ and $x^i$, or in $X$ space, where the dynamical fields are
$x^i=x^i(t,X^I)$.  The action of the model is the sum of three parts
$S=S_1+S_2+S_3$, where the first part is written in $x$ space:
\begin{equation}
  \mathcal L_1 = \frac\nu{4\pi}\epsilon^{\mu\nu\lambda} a_\mu\d_\nu a_\lambda 
  + \rho v^\mu\Bigl(\d_\mu\varphi - \tilde A_\mu - \frac12 \omega_\mu + a_\mu
   \Bigr).
\end{equation}
This has the same form as the the universal part of the
action derived in Ref.~\cite{Son:2013rqa}, but with $s$ replaced by
$1/2$.  The reason for this replacement is that we expect $L_1$ to encode
the Hall viscosity at high frequency, but not at low frequency.

The second part of the action is written in $X$ space.  It is a
Wess-Zumino-Witten action:
\begin{equation}
  S_2 = \frac{\alpha\rho_0}2\!
  \int_0^1\!d\tau\!\int\!dt\,d^2X\, \d_\tau G_{I\!J}\, G^{J\!K}\, \d_t G_{K\!L}\, \epsilon^{LI},
\end{equation}
where as a function of $\tau$, $G_{I\!J}(0,t,X)=\delta_{I\!J}$ and
$G_{I\!J}(1,t,X)=G_{I\!J}(t,X)$, and $\alpha$ is a parameter that will
be fixed later.  Although the action is written as an
integral in $\tau$ space, one can check that the action
depends only on the boundary value at $\tau=1$, but is independent of the interpolation between $\tau=0$ and $\tau=1$.  It is identical to the action considered in Ref.~\cite{Sondhi}.

Finally, in $S_3$ we include the potential energy, which depends on the
density and one parameter characterizing the eccentricity of the
deformation:
\begin{equation}
  \mathcal L_3 = \mathcal L_3( \varepsilon^{ij}\epsilon_{IJ}\d_i X^I \d_j X^J ,
   G_{IJ} h^{ij}\d_i X^I \d_j X^J ).
\end{equation}
We only consider small perturbations around the ground state; we take:
$X^I = x^I-u^I$ and  $G_{IJ}=\delta_{IJ}+H_{IJ}$.  Ignoring the constant term, total
derivatives and terms proportional to squares of $\d_i u^i$ and $h_{ii}$, which
are small in the regime we are considering, we have:
\begin{equation}
  \mathcal L_3 = -\frac{\alpha\rho_0\Delta}4 
  (\d_i u_j + \d_j u_i + h_{ij} - H_{ij} 
   )^2.
\end{equation}
Now we introduce the variable $\gamma_{ij}$ as:
\begin{equation}
  \gamma_{ij} = \d_i u_j + \d_j u_i + h_{ij} - H_{ij} .
\end{equation}
Using $\gamma$ we can rewrite the quadratic action in the form 
of eq.~\eqref{L-NC} where $\mathcal{L}_0$ is given by:
\begin{equation}\label{L-gamma2}
  \mathcal L_0 =  \frac{\alpha\rho_0}2 \left(  
    \tilde\sigma_{ij}\gamma_{ij} + \frac12 \tilde\gamma_{ij}\dot\gamma_{ij}
    - \frac\Delta 2 \gamma_{ij}^2
  \right).
\end{equation}
After integrating out $\gamma_{ij}$, $\mathcal L_0$ reduces to the
form~(\ref{L0-FG}), with functions $F$ and $G$ given by:
\begin{equation}\label{F-G}
  F(\omega) = \frac{\alpha \Delta}{\omega^2-\Delta^2+i\epsilon}\,,\quad
  G(\omega) = \frac\alpha{\omega^2-\Delta^2+i\epsilon} \, .
\end{equation}
This corresponds exactly to the spectral functions $\rho_T(\omega)\sim
\delta(\omega-\Delta)$ and $\bar\rho_T=0$.  In particular,
$\alpha=(\mathcal S-1)/4$ and the inequality~(\ref{S4-lowerbound})
becomes an equality in this model.


\subsection{Dispersion relation for the magneto-roton}

We reiterate the form of the effective Lagrangian in flat space-time and in the massless limit:
\begin{multline}\label{L-Eff}
\mathcal L = \frac\nu{4\pi}\varepsilon^{\mu\nu\lambda} a_\mu\d_\nu
  a_\lambda - \rho v^\mu(\d_\mu\varphi- A_\mu  +a_\mu)+\rho\frac{s-1}{2}\epsilon^{ij}\partial_i v_j\\
  -   \frac\rho 4\left( \sigma_{ij} F(\omega) \sigma_{ij}
  + \tilde \sigma_{ij} G(\omega) \dot\sigma_{ij} \right) -\epsilon_i(\rho),
\end{multline}
where $F$ and $G$ are given in eq. \eqref{F-G} and the function $\epsilon_i(\rho)$ denotes interaction energy of the Hall state and depends only on the particle density. 

We are interested in the dispersion relation of the magneto-roton excitations. To this end, we linearize the equations of motion and turn on perturbations about the  ground state:
\begin{align}
a_\mu=A_\mu+\tilde{a}_\mu,\quad
\rho=\rho_0+\tilde{\rho},\quad
v_i=0+v_i,
\end{align}
where $\rho_0=\frac{\nu}{2\pi}\epsilon^{ij}\partial_i A_j=\frac{\nu}{2\pi}B$ is the ground state electron density and $\tilde{a}_\mu,\tilde{\rho},v_i$ are small perturbations. We further assume that the energy arises purely from pairwise interactions with the form: 
\begin{equation}
\epsilon_i(\rho)=\frac{1}{2}\int{\int{d^2x d^2y (\rho(x)-\rho_0)V(|\mathbf{x}-\mathbf{y}|)(\rho(y)-\rho_0)}}.
\end{equation}
For the sake of definiteness, we work with the Coulomb potential with a strength parameter $\lambda$ defined as:
\begin{equation}
\frac{\lambda}{q}=\int{d^2(\mathbf{x}-\mathbf{y})V_c(|\mathbf{x}-\mathbf{y}|)e^{i(\mathbf{q(x-y)})}}.
\end{equation}
It should be noted that even though the specifics of the calculation depend on the exact form of the chosen potential, the qualitative behavior that we derive here is independent of such details, so long as the potential remains repulsive. 

In what follows we set $p=k l_B = k/\sqrt{B}$ and $\sigma_s=\frac12 (1-s)$. We note that the parameter $\alpha$ used in the non-universal functions $F$ and $G$ (see eq. \eqref{F-G}) is fixed by our first sum rule given in eq. \eqref{sbar}: $\alpha=2S_4$. Also, as noted previously, the inequality 
\eqref{S4-lowerbound} is saturated in this model.

\begin{figure}[ht!]
\centering
\includegraphics[scale=0.8]{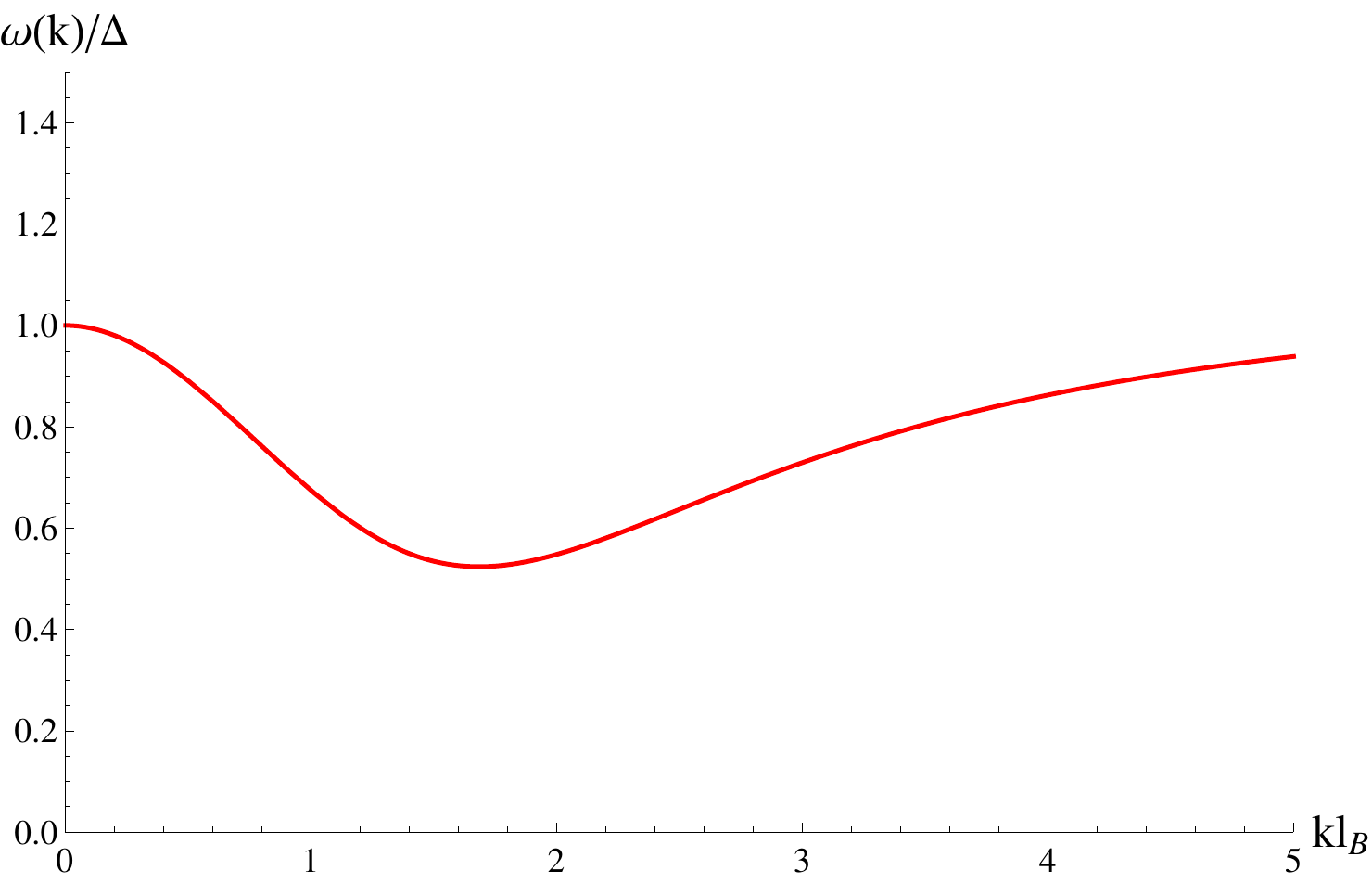}
\caption[]{Dispersion relation of the collective mode with $\nu=1/3$ and $\lambda/ (l_B\Delta)=0.3$. Minimum dispersion appears at 
$k_{min} \approx 1.69 \,l^{-1}_B  $.}
\label{Dis}
\end{figure}

We find the dispersion relation of the magneto-roton mode to be:
\begin{equation}\label{Dispersion1}
\omega(p)=\Delta\frac{\sqrt{1+2\sigma_s p^2+\dfrac{\alpha \lambda}{\Delta l_B}p^3+\sigma_s^2p^4}}{1+\big(\alpha +\sigma_s\big)p^2} \, ,
\end{equation}
which exhibits the properties of the magneto-roton with a downward slope at small $p$ which turns around after a characteristic minimum (Fig \ref{Dis}).

For the purpose of comparison, we also report this this dispersion relation up to fourth order in momentum expansion:
\begin{equation}\label{Dispersion2}
\omega(p)=\Delta\left[1-2S_4 p^2+\frac{\lambda S_4}{l_B\Delta}p^3+(4S^2_4+2S_4\sigma_s)p^4+\mathcal{O}(p^5)\right].
\end{equation}
Note that the coefficient in front of $p^2$ is, for Laughlin's fractions, $-(1-\nu)/4\nu$, which is also seen in the model of Ref.~\cite{Tokatly:2006}.
However, we cannot argue that this coefficient is universal.  For example,
the coefficient will change if we add to the action~(\ref{L-gamma2}) a term
proportional to the square of the spatial gradients of $\gamma_{ij}$.

There is one more interesting feature of this mode that reveals itself under closer inspection. From the equations of motion, we can derive that the eigenmodes satisfy: 
\begin{equation}
 \mathbf{p\times v}=if(p)\mathbf{p\cdot v}, \qquad
f(p)=\frac{1+p^2 \sigma_s +{\omega_p}^2p^2 G(\omega_p)}{\omega_p \, p^2 F(\omega_p)} \,,
\end{equation}
where $\mathbf{k\times v}=\epsilon^{ij}k_j v_j$ and $\omega_p=\omega(p)$ given in \eqref{Dispersion1}. 

\begin{figure}[ht!]
\centering
\includegraphics[scale=0.7]{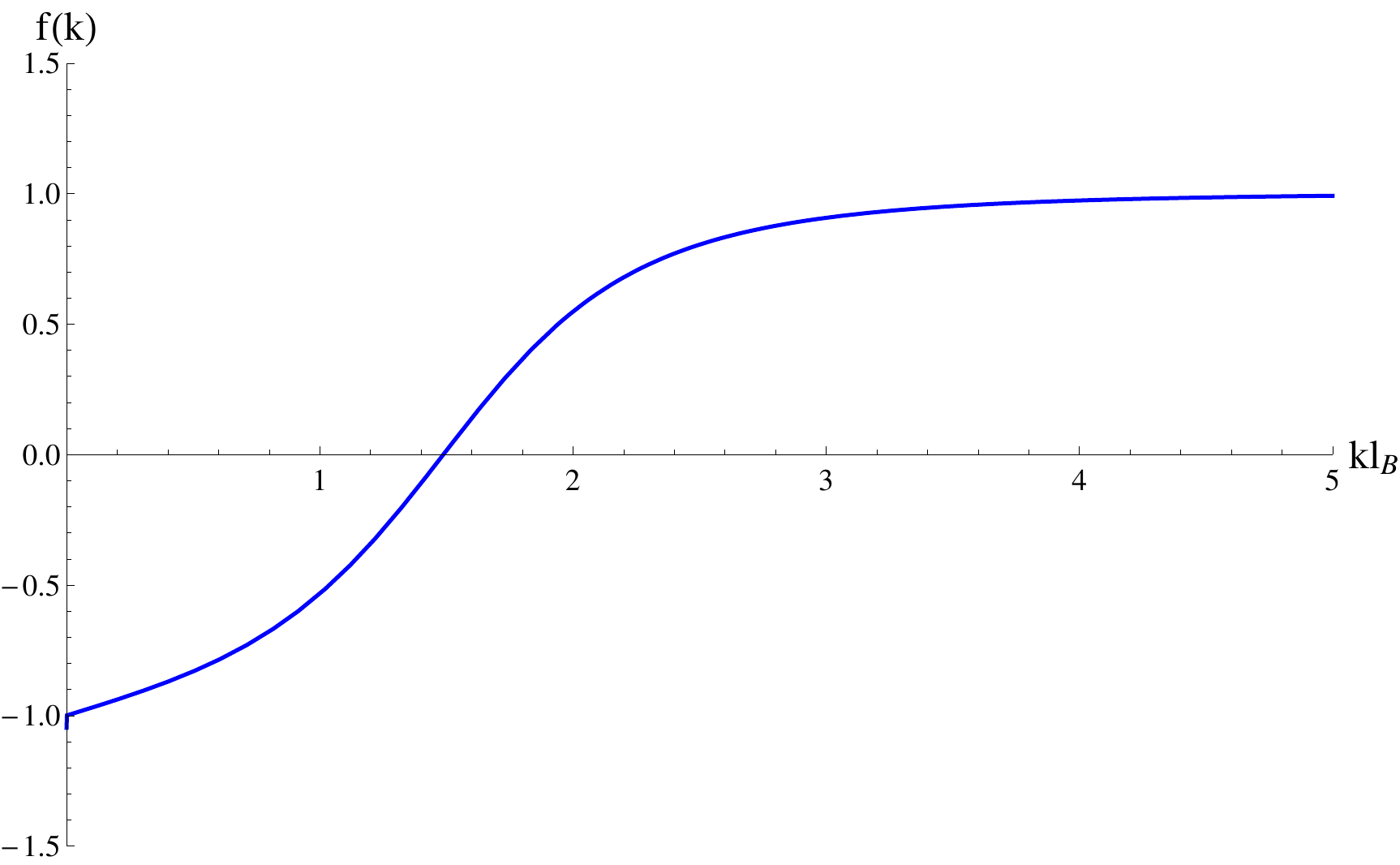}
\caption[]{Polarization function $f(k)$, with $\nu=1/3$ , $\lambda/ (l_B\Delta)=0.3$. Linear polarization appears at $k\approx 1.49 l^{-1}_B$.}
\label{Pol}
\end{figure}

If we decompose $\mathbf{v}$ into a parallel component $v_\|$ and a perpendicular component $v_\bot$ to the direction of momentum $\mathbf{p}$, we find that $v_\bot=if(p)v_\|$. Using the dispersion relation~\eqref{Dispersion1}, and the explicit form of $G(\omega),F(\omega)$, we see that the value of $f(k)$ evolves from $f(0)=-1$  to $f(k\approx k_{min})=0$ and finally to $f(k=\infty)=1$ (Fig \ref{Pol}).  

This implies that, from the point of view of current pattern, the excitations exhibit counterclockwise rotation at
small momenta, which turns into a linear oscillation in direction of
$\mathbf p$ in the vicinity of the magneto-roton minimum and finally
develops into clockwise rotation at large values of the momentum.  It would
be interesting to understand if this
feature of the magneto-roton may be detected experimentally.

Here we only briefly discuss the implication of the model for the
observation of the magneto-roton at low momentum in Raman scattering
experiments.  In previous theoretical treatments~\cite{PlatzmanHe}, it
was assumed that the magneto-roton is excited chiefly through the
coupling of electric field to density: $\rho E^2$.  This coupling,
however, implies that the intensity of the magneto-roton peak should
scale as the fourth power of the magneto-roton momentum, $q^4$.  In
experiments, magneto-roton was seen down to even the lowest
momenta~\cite{Pinczuk}, a fact that may be attributed to disorders
violating momentum conservation.  However, we cannot rule out a
coupling of the type $T_{ij}E_i E_j$ from symmetry consideration, with
$T_{ij}$ being the stress tensor.  Even if the coefficient in front of
this term is small, it would dominate the intensity of magneto-roton
peak in the limit $q\to0$, since the residue at the pole in in
$\<TT\>$ correlators remains finite in this limit.  This coupling thus
provides an alternative explanation of the observation of the
magneto-roton at lowest momenta in Raman scattering experiments.

Moreover, in our model at $q=0$ the magneto-roton is circularly
polarized with angular momentum 2.  We suggest that the
polarization of the magneto-roton at $q=0$ may be detectable by Raman
scattering with polarized light.

\section{Conclusion}
In this paper we looked at gapped fractional quantum Hall states with filling factors $\nu<1$ in the regime where the Coulomb energy is much smaller than
the cyclotron energy, however with energies comparable to that of the gap. We developed three sum rules involving the spectral densities of the stress tensor which we then used to verify Haldane's conjectured lower bound on the quartic coefficient of the structure factor $\mathcal S_4$, as well as introduce other inequalities.

We also introduced a simple model that saturates these inequalities
via a mode which arises from the mixing between the 
oscillations of an internal metric and the hydrodynamic excitations.
We identifed this mode as the magneto-roton and calculated its
dispersion relation.  We argued that the intensity of the magneto-roton
line in Raman scattering experiments should not vanish at zero
momentum, and that the magneto-roton at $q=0$ is a spin-2 object.
Finally, we suggest that the spin of the magneto-roton can be determined
by Raman scattering with polarized light.

\acknowledgments 

The authors thank Ilya Gruzberg, Michael Levin, Emil Martinec, Aron
Pinczuk, and Paul Wiegmann for discussion.  This work is supported, in
part, by NSF MRSEC grant DMR-0820054.  D.T.S. is supported, in part,
by DOE grant DE-FG02-90ER-40560 and a Simons Investigator grant from
the Simons Foundation.

\end{document}